\documentclass{ifacconf}

\usepackage{amsmath}
\usepackage{amssymb}  
\usepackage{algorithm}
\usepackage{algpseudocode}
\usepackage[utf8]{inputenc}
\usepackage{cancel}

\usepackage{graphicx}      
\usepackage{epsfig}
\usepackage{natbib}        
\begin{document}
\begin{frontmatter}

\title{Adaptive Observers for MIMO Discrete-Time LTI Systems} 


\author[First]{Anchita Dey} 
\author[Second]{Shubhendu Bhasin} 
\address[First]{Electrical Engineering Department, Indian Institute of Technology Delhi, New Delhi 110016, India (e-mail: anchita.dey@ee.iitd.ac.in)}
\address[Second]{Electrical Engineering Department, Indian Institute of Technology Delhi, New Delhi 110016, India (e-mail: sbhasin@ee.iitd.ac.in)}

\begin{abstract}
In this paper, an adaptive observer is proposed for multi-input multi-output (MIMO) discrete-time linear time-invariant (LTI) systems. Unlike existing MIMO adaptive observer designs, the proposed approach is applicable to LTI systems in their general form. Further, the proposed method uses recursive least square (RLS) with covariance resetting for adaptation that is shown to guarantee that the estimates are bounded, irrespective of any excitation condition, even in the presence of a vanishing perturbation term in the error used for updation in RLS. Detailed analysis for convergence and boundedness has been provided along with simulation results for illustrating the performance of the developed theory.            
\end{abstract}

\begin{keyword}
Adaptive observer design, Recursive identification, Linear multivariable systems, Observer design, Observers for linear systems.
\end{keyword}

\end{frontmatter}

\section{Introduction}
Many control applications rely on the availability of accurate system models and state measurements. The fact that such information is often difficult or expensive to obtain precludes many controllers from wider application. Several adaptive and robust designs in the literature address the problem either partially or entirely, i.e., by considering the unavailability of either an exact model or perfect state information, or by taking into account the unavailability of both. Adaptive observers provide one approach to solve the problem by simultaneously estimating the state and system parameters online. 

Majority of the adaptive observers in literature are designed for continuous-time systems using a Lyapunov synthesis technique, e.g., \cite{carroll1973adaptive} and \cite{luders1974new}. The convergence result in \cite{carroll1973adaptive} requires an additional auxiliary signal, a requirement that is relaxed in \cite{luders1974new} by using a new canonical realization for the observer. A significant improvement in observer design is made in \cite{kreisselmeier1977adaptive} where the parameter update loop is separated from the observer dynamics used to find the state estimate; this allows different designs of the update law without disturbing the adaptive observer dynamics. Multi-input multi-output (MIMO) continuous-time adaptive observers are proposed in \cite{anderson1974adaptive} and \cite{zhang2002adaptive}. \cite{anderson1974adaptive} uses a transfer function based approach, whereas \cite{zhang2002adaptive} considers the dynamics of the linear system in a specific form where the state and input are multiplied with known parameters and an additive term equal to the product of an unknown parameter and a known regressor is present. The known regressor is dependent on only input and output data, and it is not possible to cast this model into the standard form of linear time-invariant (LTI) models. For continuous-time nonlinear systems with single output, an adaptive observer that is robust to bounded disturbance is provided in \cite{marine2001robust}, where a projection operator is used to prevent parameter drift due to the disturbances.

In contrast to their continuous-time counterparts, few adaptive observers exist for discrete-time systems; some notable works include \cite{kudva1974discrete}, \cite{tamaki1981design} and \cite{suzuki1980discrete} where the first two use Lyapunov's direct method to synthesize the adaptive observer. The philosophy in \cite{kreisselmeier1977adaptive} has been utilized in \cite{suzuki1977design} to design an adaptive observer that uses exponential forgetting factor based recursive least square (RLS) for learning the system parameters of the discrete-time LTI system.

The above-mentioned works on adaptive observer designs have mostly been carried out for only single-input single-output (SISO) systems, and few notable works that exist for the MIMO cases are only available for continuous-time systems. Additionally, all of these works depend on a persistently exciting (PE) regressor or a sufficiently rich input for guaranteeing bounded estimates, for e.g., \cite{zhang2002adaptive}. Relaxation of the excitation condition is proposed in some recent works for continuous-time systems, e.g., \cite{katiyar2022initial}, that is again based on the system structure used in \cite{zhang2002adaptive}, and \cite{tomei2022enhanced}, that is designed for the SISO case. A different approach using set membership based estimation of state and parameters is given in \cite{pan2022set}; however, the method is computationally heavy and requires initial knowledge of a set where the state and parameters belong.

In this paper, an adaptive observer for MIMO discrete-time LTI systems is proposed. To the best of the authors' knowledge, this is the first result on discrete-time MIMO adaptive observers that is applicable to the general form of LTI systems, as compared to the result in \cite{zhang2002adaptive} that holds for a less general case of LTI systems. Further, the adaptive observer is shown to  generate bounded state and parameter estimates even in the absence of any excitation condition. Similar to \cite{suzuki1980discrete}, the state and output vectors are written in a linear regressor form using filtered variables. However, the design of the filtered variables is challenging owing to the presence of multiple inputs and outputs. Also, the direct extension of the SISO design in \cite{suzuki1980discrete} leads to repeated parameters in the parameter vector. Hence, a modification using Kronecker product has been done to design suitable filtered variables that avoids repetition in estimation of any parameter.

For parameter learning, a point-based estimation using RLS is chosen to reduce computation and enable online identification. However, ordinary RLS (\cite{goodwin1984adaptive}) results in slow convergence after the covariance matrix becomes small; on the other hand, the exponential forgetting factor based RLS (\cite{suzuki1980discrete}) leads to parameter drift and consequently, unbounded state estimates, in absence of sufficient excitation. The latter can be prevented using variable rate forgetting factor but with an additional law for bounding the covariance as proposed in \cite{shah1991recursive}. Alternatively, the covariance may be reset whenever required to ensure better convergence as suggested in \cite{goodwin1984adaptive}. Several variants of resetting the covariance for better convergence given in \cite{tham1988covariance} depend on the excitation condition. The improved least square algorithm in \cite{rao1987improved} carries out no adaptation whenever the input excitation is insufficient by additionally checking the latter as well. {In this paper, the covariance is reset by checking its minimum eigenvalue. The resetting does not allow the minimum eigenvalue to become infinitesimally small, which otherwise, could have led to extremely slow parameter convergence along the eigenvectors corresponding to the infinitesimally small eigenvalues of the covariance. The adaptation is never halted unlike the method in \cite{rao1987improved}. Placing a lower bound on the minimum eigenvalue and resetting the covariance back to its initial value ensures that the covariance is both upper and lower bounded. The upper bound is needed to prevent covariance windup (\cite{rao1987improved}), whereas the lower bound ensures that the learning is not inhibited due to insignificant covariance.}

An additional difficulty is due to the initial state estimation error which appears as a bounded but exponentially decaying term in the linear regressor form of the output equation. A similar term appears in \cite{kreisselmeier1977adaptive} where the detailed analysis for the transient part is not provided. Deadzone RLS as mentioned in \cite{lozano1987reformulation} can be used in the presence of such perturbation; however, this leads to extra computation owing to the set-based analysis. Interestingly, it is possible to use RLS with covariance resetting to guarantee that the parameter and state estimates are always bounded and robust to the vanishing perturbation due to the initial state estimation error. A detailed analysis of boundedness is provided in this paper.

 The contribution of the paper is the design of the discrete-time adaptive observer for MIMO LTI systems in its general form, using the covariance resetting based RLS technique with thorough analysis of boundedness and convergence of the estimates, irrespective of the PE condition.

\subsubsection{Notations:}
$I_q\in\mathbb{R}^{q\times q}$ and $0_q\in\mathbb{R}^{q\times q}$ denote the Identity and zero matrix, respectively. $\mathbb{I}_q^+$ is the set of all integers from $q$ to $\infty$, and $\mathbb{I}_q^r\triangleq \{q,\;q+1,\;...,\;r-1,\;r\}$ where $q<r$. For a matrix $M=\begin{bmatrix}M_1&M_2&...&M_r\end{bmatrix}\in\mathbb{R}^{q\times r}$ where $M_i\in\mathbb{R}^q$ $\forall i\in\mathbb{I}_1^r$, $vec(M)\triangleq \begin{bmatrix}M_1^\intercal&M_2^\intercal&...&M_r^\intercal\end{bmatrix}^\intercal\in\mathbb{R}^{qr}$. $\lambda[\cdot]$, $\lambda_{min}[\cdot]$ and $\lambda_{max}[\cdot]$ denote eigenvalue, the minimum eigenvalue and the maximum eigenvalue of a matrix, respectively, $\otimes$ denotes the Kronecker product, and $||\cdot||$ represents the Euclidean (induced Euclidean) norm of a vector (matrix). For two matrices $M$ and $N$, $M\succ (\succeq) N$ or $N \prec (\preceq) M$ implies $M-N$ is positive-definite (positive semi-definite). Any signal $g_t\in\mathcal{L}_\infty$ denotes that $g_t$ is bounded.

\section{Problem Formulation}
Consider an unknown MIMO LTI system with transfer function matrix, after all pole-zero cancellations, as a strictly proper rational matrix $G_{sp}(s)$ given by
\begin{equation}\label{Gsp}
    G_{sp}(s)=\frac{\textbf{N}_1s^{r-1}+\textbf{N}_2s^{r-2}+...+\textbf{N}_{r-1}s+\textbf{N}_r}{s^r+a_1s^{r-1}+...+a_{r-1}s+a_r}
\end{equation}
where $r\in\mathbb{I}_1^+$, and $\textbf{N}_i\in\mathbb{R}^{q\times m}$, $a_i\in\mathbb{R}$ $\forall i\in\mathbb{I}_1^r$. Since the system is unknown, the values of $\textbf{N}_i$ and $a_i$ in \eqref{Gsp} are unavailable. The primary goal of the paper is to simultaneously identify online the parameters $\textbf{N}_i$ and $a_i$.

It is possible to construct a realization of \eqref{Gsp} in the observable canonical form 
\begin{equation}\label{sys}
    x_{t+1}=Ax_t+Bu_t\text{ ; }\;\;\;\; y_t=Cx_t
\end{equation}
where $u_t\in\mathbb{R}^m$ denotes the input, $y_t\in\mathbb{R}^q$ denotes the output, $x_t\in\mathbb{R}^n$ (where, ${n\triangleq rq}$ and $n\in\mathbb{I}_1^+$) is the state, and $A\in\mathbb{R}^{n\times n}$, $B\in\mathbb{R}^{n\times m}$ and $C\in\mathbb{R}^{q\times n}$ are the system matrices. In observable canonical form, the system matrices can be written as \cite[Ch.~4]{chen1984linear}
\begin{align}
A=   & \begin{bmatrix}
   -a_1I_q&& I_q &0_q& ... &0_q \\
   -a_2I_q && 0_q &I_q & ... &0_q\\
   \vdots & &\vdots &\vdots & \ddots &\vdots\\
   -a_{r-1}I_q& & 0_q & 0_q & ... & I_q\\
   -a_rI_q && 0_q & 0_q & ... & 0_q
    \end{bmatrix} 
    \text{, }
    B=\begin{bmatrix}\textbf{N}_1\\ \textbf{N}_2\\  \vdots\\ \textbf{N}_r
\end{bmatrix}\label{ab}
\\
\text{and }C =&\begin{bmatrix}
    I_q & 0_q & ... & 0_q
    \end{bmatrix}.\label{c}
\end{align}
The \textit{objective} reduces to estimating the unknown parameters $A$ and $B$, along with the unavailable state measurements $x_t$. Motivated by \cite{kreisselmeier1977adaptive} and \cite{suzuki1980discrete}, an adaptive observer is designed to achieve the objective, however, unlike \cite{kreisselmeier1977adaptive} and \cite{suzuki1980discrete}, where the observer is designed for only SISO LTI systems, the proposed method is applicable to the MIMO case. Further, the proposed adaptive law guarantees boundedness of the state and parameter estimates, irrespective of the excitation of the regressor, unlike the designs in \cite{kreisselmeier1977adaptive} and \cite{suzuki1980discrete} which do not provide performance guarantees in absence of a PE regressor.

\section{MIMO Adaptive Observer}
The matrix $B$ in \eqref{ab} is rewritten as 
\begin{equation}
    B=\begin{bmatrix}
  b_{11} & b_{12} & ... & b_{1m}\\
  \vdots & \vdots& \ddots & \vdots \\
  b_{n1}&b_{n2}&...&b_{nm}
  \end{bmatrix}
\end{equation}
where $b_{ij}$ $\forall (i,j)\in\mathbb{I}_1^n\times \mathbb{I}_1^m$ are scalars. Let $F$ be a Schur stable matrix structurally similar to $A$ where
\begin{align}\label{F}
    F\triangleq\begin{bmatrix}
     -f_1I_q& & I_q &0_q& ... &0_q \\
   -f_2I_q & & 0_q &I_q & ... &0_q\\
   \vdots & & \vdots &\vdots & \ddots &\vdots\\
   -f_{r-1}I_q & & 0_q & 0_q & ... & I_q\\
   -f_rI_q & & 0_q & 0_q & ... & 0_q
    \end{bmatrix}\in\mathbb{R}^{n\times n}.
\end{align}
All the unknown parameters in \eqref{ab} are assembled in two vectors $a\triangleq-\begin{bmatrix}a_1 & a_2 & ... &a_r
  \end{bmatrix}^\intercal  \in\mathbb{R}^r\text{ and }b\triangleq vec{(B)} \in\mathbb{R}^{nm}$. Also, let $f\triangleq -\begin{bmatrix}f_1 & f_2& ... & f_r\end{bmatrix}^\intercal\in\mathbb{R}^r$. Using \eqref{F}, $a$ and $f$, \eqref{sys} can be re-written as
\begin{equation}\label{sysF}
    x_{t+1}=Fx_t+Y_t(a-f)+Bu_t
\end{equation}
where $Y_t\triangleq I_r\otimes y_t\in\mathbb{R}^{n\times r}$. Let $S_{y_t}\in\mathbb{R}^{n\times r}$ and $S_{u_t}^{(i)}\in\mathbb{R}^{n\times n}$ $\forall$ $i\in\mathbb{I}_1^m$ be $m+1$ filter matrices that satisfy
\begin{align}
   & S_{y_{t+1}}=F S_{y_t}+Y_t\text{ ; }S_{y_0}=0_{n\times r} \label{S1}\\
  &  S_{u_{t+1}}^{(i)}=FS_{u_t}^{(i)}+I_n u_t^{(i)}\text{ ; }S_{u_0}^{(i)}=0_{n\times n}\hspace{0.3cm}\forall\text{ }i\in\mathbb{I}_1^m \label{S2}
  \end{align}where $u_t^{(i)}$ denotes the $i^{\text{th}}$ element in $u_t$. Further,
  \begin{align}
  &  S_{u_t}\triangleq\begin{bmatrix}S_{u_t}^{(1)}& S_{u_t}^{(2)} & ... & S^{(m)}_{u_t}\end{bmatrix}\in\mathbb{R}^{n \times mn} \label{S3}\\
  \text{and } & S_t\triangleq\begin{bmatrix}S_{y_t}&S_{u_t}\end{bmatrix}\in\mathbb{R}^{n\times (r+mn) }\label{S4}
\end{align}

Using the unknown parameter vector $p\triangleq\begin{bmatrix}(a-f)^\intercal  & b^\intercal  \end{bmatrix}^\intercal \in\mathbb{R}^{r+mn}$, its estimate $\hat{p}_t\triangleq\begin{bmatrix}(\hat{a}_t-f)^\intercal  & \hat{b}_t^\intercal  \end{bmatrix}^\intercal \in\mathbb{R}^{r+mn}$ (with $\hat{a}_t$ and $\hat{b}_t$ being the estimates of $a$ and $b$, respectively), and \eqref{S1}-\eqref{S4}, the solution of \eqref{sysF} and the output are respectively given by
\begin{equation}\label{xxt}
    {x}_t=S_tp+F^t{x}_0\hspace{0.2cm}\text{ and }\hspace{0.2cm} {y}_t=\phi^\intercal _tp+CF^t{x}_0
\end{equation}where $\phi_t\triangleq S^\intercal _tC^\intercal \in\mathbb{R}^{(r+mn)\times q}$. Using \eqref{xxt}, an adaptive observer is designed, with the state and output given by
\begin{align}\label{xhat}
    \hat{x}_t=S_t\hat{p}_t+F^t\hat{x}_0 \text{ and }\hat{y}_t=\phi^\intercal _t\hat{p}_t+CF^t\hat{x}_0.
\end{align}
 In addition, from \eqref{xxt} and \eqref{xhat}, the state estimation error $\tilde{x}_t\triangleq x_t-\hat{x}_t\in\mathbb{R}^n$ is given by
\begin{equation}\label{xtil}
    \tilde{x}_t=S_t\tilde{p}_t+F^t\tilde{x}_0
\end{equation}where $\tilde{p}_t\triangleq p-\hat{p}_t\in\mathbb{R}^{r+mn}$ is the parameter estimation error. Clearly, the second term in \eqref{xhat} is exponentially decaying to zero; thus, the convergence of $\tilde{x}_t$ to zero can be guaranteed by designing a suitable parameter estimator that ensures the convergence of $\tilde{p}_t$ to zero. 

Further, \eqref{xhat} implies that the state is estimated using the current parameter estimate. At each time step, the parameter estimates are computed first, using a subsequently derived adaptation law, which are then used for state estimation. 
\begin{defn}\label{PEdef}
The regressor $\phi_t$ is said to be PE if  \cite[Eq.~3.4.26]{goodwin1984adaptive}
    \begin{align}\label{PE}
        \lim_{t\rightarrow\infty}\lambda_{min}\Big{[}\sum_{i=1}^t \phi_i\phi_i^\intercal\Big{]}=\infty.
    \end{align}
\end{defn}


\subsection{Design of the Parameter Estimation Law}
An RLS law is designed that updates $\hat{p}_t$ using the output estimation error defined as
\begin{equation}\label{rege}
    \tilde{y}_t\triangleq y_t-\hat{y}_t=\underbrace{y_t-CF^t\hat{x}_0}_{=z_t}-\phi^\intercal _t\hat{p}_t=z_t-\phi_t^\intercal \hat{p}_t.
\end{equation}
Minimization of the cost function \cite[eq.~(3.8.3)]{goodwin1984adaptive} \begin{gather}\begin{aligned}
    J_t(\hat{p})\triangleq\frac{1}{2}\Big{[}
    \sum_{i=1}^t\{(z_i-&\phi_i^\intercal \hat{p})^\intercal R^{-1} (z_i-\phi_i^\intercal \hat{p})\}
    \\&+(\hat{p}-\hat{p}_0)^\intercal \Gamma^{-1}_0(\hat{p}-\hat{p}_0)\Big{]}
\end{aligned}\end{gather} 
where $R\in\mathbb{R}^{q\times q}$ and $\Gamma_0\in\mathbb{R}^{(r+mn)\times(r+mn)}$ are user-defined symmetric positive-definite matrices, results in the minimizer $\hat{p}_t$ that satisfies
\begin{equation}\label{phat}  \hat{p}_t=\Gamma_t \Big{(} \Gamma^{-1}_0\hat{p}_0 + \sum_{i=1}^t\phi_iR^{-1}z_i\Big{)}
\end{equation}
where $\Gamma_t^{-1}= \Gamma_0^{-1}+\sum_{i=1}^t \phi_iR^{-1}\phi_i^\intercal $ is a positive-definite matrix. $\Gamma_t$ is the covariance matrix that satisfies the recursion
\begin{equation}\label{Gammainv}
    \Gamma_{t+1}^{-1}=\Gamma^{-1}_{t}+\phi_{t+1}R^{-1}\phi^\intercal_{t+1} \;\;\;\;\;\;\forall t\in\mathbb{I}_0^+.
\end{equation}
The RLS update laws obtained with simple manipulation on \eqref{phat} using \eqref{Gammainv} are
\begin{align}
     & \hat{p}_{t+1}=\hat{p}_{t}+\Gamma_{t+1}\phi_{t+1}R^{-1}\Big{(} z_{t+1}-\phi_{t+1}^\intercal \hat{p}_{t} \Big{)} \;\;\text{ }\forall t\in\mathbb{I}_0^+ \label{phat1}\\
   & \text{where } \Gamma_{t+1}=\Gamma_{t}-\Gamma_{t}\phi_{t+1} W_{t+1} \phi_{t+1}^\intercal \Gamma_{t-1}\label{Gamma1}
\end{align}
and ${W}_t\triangleq\Big{(} R+\phi_t^\intercal \Gamma_{t-1}\phi_t \Big{)}^{-1}\in\mathbb{R}^{q\times q}$. 
Use of the matrix inversion lemma\footnote{$(O+PQX)^{-1}=O^{-1}-O^{-1}P(Q^{-1}+XO^{-1}P)^{-1}XO^{-1}$} on \eqref{Gammainv}-\eqref{Gamma1} results in
\begin{equation}\label{phatg}
   \hat{p}_{t+1}=\hat{p}_{t}+\Gamma_{t}\phi_{t+1}W_{t+1} \Big{(} z_{t+1}-\phi_{t+1}^\intercal \hat{p}_{t} \Big{)} 
\end{equation}
which is typically used instead of \eqref{phat1} for computation of parameter estimates.

\subsubsection{Covariance resetting:}
In ordinary RLS, the covariance matrix in the update law reduces drastically in the first few iterations; this prevents proper learning of the unknown parameters. A possible solution to this problem is covariance resetting where the covariance $\Gamma_t$ is reset to some higher covariance value at specific time intervals. 

There are several ways of resetting the covariance, for instance, at some pre-decided time instants \cite[Ch.~3]{goodwin1984adaptive}, by checking the trace of $\Gamma_t$ (\cite{tham1988covariance}), or by placing lower and upper bounds on $\Gamma_t$ along with exponential resetting and forgetting (\cite{salgado1988modified}). For the estimator designed in this paper, the following algorithm is used for updating $\Gamma_t$.
\begin{gather}\label{Gammacases}\begin{aligned}
    \bar{\Gamma}_t\triangleq& \Gamma_{t-1}-\Gamma_{t-1}\phi_tW_t\phi_t^\intercal\Gamma_{t-1}\\
    \Gamma_t=&\begin{cases}
   \Gamma_0\triangleq k_0I_{r+mn}\hspace{1cm} \text{if }\lambda_{min}[\bar{\Gamma}_t]\leq k_{min}\\
   \bar{\Gamma}_t\hfill \text{ otherwise}
       \end{cases}
\end{aligned}\end{gather}
where $k_0$, $k_{min}\in\mathbb{R}$ are user-defined positive constants and $k_{min}<k_0$. The update law \eqref{Gammacases} imposes lower and upper bounds on $\Gamma_t$ such that it is positive-definite $\forall t\in\mathbb{I}_0^+$. Further, resetting based on the minimum eigenvalue ensures that learning does not become extremely slow when the covariance becomes small.

Using \eqref{phatg} and \eqref{Gammacases} makes the adaptive observer dynamics nonlinear. A complete analysis of how the parameter and state estimates vary is necessary to ensure that there is no drift in the estimates. The proof of convergence of parameter estimation error $\tilde{p}_t\triangleq p-\hat{p}_t\in\mathbb{R}^{r+mn}$ is detailed in Lemma 3.3.8 of \cite{goodwin1984adaptive} for the case when the output estimation error is in the usual linear regressor form, i.e., $\tilde{y}_t=\phi_t^\intercal \tilde{p}_t$. However, in \eqref{rege}, the output estimation error $\tilde{y}_t=\phi_t^\intercal\tilde{p}_t+CF^t\tilde{x}_0$  has the additional term $CF^t\tilde{x}_0$. The bounded perturbation $CF^t\tilde{x}_0$ prevents the direct extension of the lemma to the update law proposed in \eqref{phatg} and \eqref{Gammacases}. For such cases when a perturbation term is present, \cite{lozano1987reformulation} and \cite{goodwin1984adaptive} propose deadzone based RLS techniques, where learning is paused whenever the error $\tilde{y}_t$ becomes smaller compared to the bound of the perturbation. Contrary to their approach, parameter learning is never paused with the  proposed estimator. A detailed analysis of the transient behaviour of the estimates obtained using \eqref{phatg} and \eqref{Gammacases}, irrespective of any excitation condition on the regressor, is not available in the literature. The analysis is provided in the next section, and is based on the following lemma.
\begin{lem}\label{prop1}
A stabilizing and bounded input $u_t$ for the LTI system \eqref{sys} ensures that the regressor $\phi_t\in\mathcal{L}_\infty$ and the matrix $W_t\in\mathcal{L}_\infty$ $\forall t\in\mathbb{I}_0^+$. \end{lem}
\begin{pf}
With a stabilizing and bounded input $u_t$ for the LTI system \eqref{sys}, the state $x_t\in \mathcal{L}_\infty$ which implies that the output $ y_t\in\mathcal{L}_\infty$ $\forall t\in\mathbb{I}_0^+$. With bounded $x_t$, $u_t$ and $y_t$, it can be derived using \eqref{S1}-\eqref{S4} that the filter variable $S_t\in\mathcal{L}_\infty$ $\forall t\in\mathbb{I}_0^+$. This ensures that $\phi_t\in\mathcal{L}_\infty$ (using \eqref{S1}-\eqref{S4} and definition of $\phi_t$). From \eqref{Gammacases}, $\lambda_{max}[\Gamma_t]$ is upper bounded by $k_0$ and $\lambda_{min}[\Gamma_t]$ is lower bounded by $k_{min}$ $\forall t\in\mathbb{I}_0^+$. Therefore, the bounded covariance $\Gamma_t$ and regressor $\phi_t$ leads to $W_t\in\mathcal{L}_\infty$ $\forall t\in\mathbb{I}_0^+$. 
\hfill$\blacksquare$
\end{pf}

\section{Convergence and Boundedness}
From \eqref{rege} and \eqref{phatg}, the parameter estimation error dynamics is given by
\begin{equation}\label{ptil1}
    \tilde{p}_{t+1}=\tilde{p}_{t}-\Gamma_t\phi_{t+1} W_{t+1}(\phi_{t+1}^\intercal\tilde{p}_{t}+CF^{t+1}\tilde{x}_0)\text{ }\;\forall t\in\mathbb{I}_0^+
\end{equation}
where $\Gamma_t$ is obtained from the covariance resetting algorithm in \eqref{Gammacases}.  The boundedness analysis $\forall t\in\mathbb{I}_0^+$ is given in Theorem \ref{allb} and the convergence analysis at steady state is given in Theorem \ref{infconv}. 
\begin{lem}\label{a0}
The matrix $\phi_{t}R^{-1}\phi_{t}^\intercal\succeq 0_{r+mn}$ $\;\forall t\in\mathbb{I}_0^+$. In addition, if $\phi_t$ is PE, then $\lim_{t\rightarrow\infty}\phi_{t}R^{-1}\phi_{t}^\intercal\succ 0_{r+mn}$.
\end{lem}
\begin{pf}
Refer to Appendix \ref{Append0}.\hfill $\blacksquare$
\end{pf}

\begin{lem}\label{a1}
For $\Gamma_t\succ 0_{r+mn}$ and $\phi_{t+1}R^{-1}\phi_{t+1}^\intercal \succeq 0_{r+mn}$, 
\begin{enumerate}
    \item $M_{1_{t+1}}\triangleq \Gamma_t\phi_{t+1}R^{-1}\phi_{t+1}^\intercal$ is similar to a positive semi-definite matrix $M_{2_{t+1}} \triangleq \Gamma_t^{\frac{1}{2}}\phi_{t+1}R^{-1}\phi_{t+1}^\intercal \Gamma_t^{\frac{1}{2}}$, and
    \item $(I_{r+mn}+M_{1_{t+1}})^{-1} \preceq I_{r+mn}$.
\end{enumerate} 
\end{lem}
\begin{pf}
 (1) Refer to \cite[Theorem~2.2]{wu1988products}. (2) Refer to Appendix \ref{Append1}.\hfill $\blacksquare$
\end{pf}

\begin{thm}\label{allb}
If the input $u_t$ to \eqref{sys} and \eqref{S2} is stabilizing and bounded, then from Lemma \ref{prop1}, \ref{a0} and \ref{a1}, the parameter estimation error $\tilde{p}_t$ and the state estimation error $\tilde{x}_t$ are bounded $\forall t\in\mathbb{I}_0^+$.
\end{thm}
\begin{pf}
The following expression for $\tilde{p}_{t+1}$ is obtained using the matrix inversion lemma on \eqref{ptil1}.
\begin{gather}\begin{aligned}\label{ptil2}
    \tilde{p}_{t+1}=(I_{r+mn}&+\Gamma_t\phi_{t+1}R^{-1}\phi_{t+1}^\intercal)^{-1}\tilde{p}_{t}\\
    &- \Gamma_t\phi_{t+1}W_{t+1}CF^{t+1}\tilde{x}_0\;\;\; \;\;\forall t\in\mathbb{I}_0^+.
\end{aligned}\end{gather}
Using Lemma \ref{a1}, (2) and \eqref{ptil2} leads to
\begin{align}\label{ptil3}
   || \tilde{p}_{t+1}||\leq ||\tilde{p}_{t}||+ ||\Gamma_t\phi_{t+1}W_{t+1}CF^{t+1}\tilde{x}_0||\; \;\;\;\forall t\in\mathbb{I}_0^+.
\end{align}
Following Lemma \ref{prop1}, $||\phi_t||\leq \beta_\phi$ and $||W_t||\leq \beta_W$ where $\beta_\phi,\;\beta_W\in\mathbb{R}$ are some finite positive constants. Using these bounds and \eqref{Gammacases}, \eqref{ptil3} can be further upper bounded as
\begin{align}
& ||\tilde{p}_{t+1}|| \leq ||\tilde{p}_{t}|| +k_0\beta_\phi\beta_W||CF^{t+1}\tilde{x}_0|| \nonumber\\
 \Rightarrow  &||\tilde{p}_{t+1}|| \leq ||\tilde{p}_{0}|| +\sum_{i=0}^t k_0\beta_\phi\beta_W||CF^{i+1}\tilde{x}_0||\;\;\;\; \forall t\in\mathbb{I}_0^+. \label{psum1}
\end{align}
Since $F$ is Schur stable, the infinite summation term ${\beta_{\infty}} \triangleq$ $\lim_{t\rightarrow\infty}{\sum_{i=0}^t {k_0\beta_\phi \beta_W ||CF^{i+1}\tilde{x}_0||}}$ is finite. Therefore, $||\tilde{p}_t||$ can be bounded as
\begin{align}\label{ultb}
    ||\tilde{p}_{t+1}|| \leq ||\tilde{p}_{0}|| +\beta_\infty \; \;\;\;\forall t\in\mathbb{I}_0^+.
\end{align}
With a finite $\hat{p}_0$, the term $||\tilde{p}_0||$ in \eqref{ultb} is finite, and as a result, $\tilde{p}_t\in\mathcal{L}_\infty\Rightarrow \hat{p}_t\in\mathcal{L}_\infty$ $\;\forall t\in\mathbb{I}_0^+$. Using this and Lemma \ref{prop1}, from \eqref{xtil}, $\tilde{x}_t\in\mathcal{L}_\infty \Rightarrow \hat{x}_t\in\mathcal{L}_\infty$ $\;\forall t\in\mathbb{I}_0^+$.\hfill $\blacksquare$ \end{pf}

\begin{thm}\label{infconv}
With a stabilizing and bounded input $u_t$ applied to \eqref{sys} and \eqref{S2}, and using Lemma \ref{prop1}, \ref{a0} and \ref{a1}, the parameter estimation error $\tilde{p}_t$ is guaranteed to converge to some finite value as $t\rightarrow\infty$. Additionally, if the regressor $\phi_t$ is PE (Definition \ref{PEdef}), then $\tilde{p}_t$ converges to zero as $t\rightarrow\infty$. Further, if $\tilde{p}_t$ converges to zero as $t\rightarrow\infty$, then the state estimation error $\tilde{x}_t$ also converges to zero as $t\rightarrow\infty$.
\end{thm}
\begin{pf}
In \eqref{ptil2}, the second term vanishes exponentially as $t\rightarrow\infty$. The convergence analysis is carried out at $t\rightarrow\infty$ using the Lyapunov function $V_t\triangleq\tilde{p}_t^\intercal \Gamma_t^{-1}\tilde{p}_t$ where $\tilde{p}_t$ obeys the dynamics
\begin{align}
     \tilde{p}_{t+1}=\tilde{p}_{t}-\Gamma_t\phi_{t+1} W_{t+1}\phi_{t+1}^\intercal\tilde{p}_{t}.
\end{align}
Without the vanishing perturbation term $CF^t\tilde{x}_0$, the remaining proof for convergence of parameter estimation error $\tilde{p}_t$ directly follows Lemma 3.3.8 of \cite{goodwin1984adaptive}. 

In \eqref{xtil}, the second term is exponentially decaying since $F$ is Schur stable. Therefore, if $\tilde{p}_t$ converges to zero as $t\rightarrow\infty$, then from \eqref{xtil}, $\tilde{x}_t$ also converges to zero as $t\rightarrow\infty$.\hfill$\blacksquare$


\end{pf}


\section{A Numerical Example}
We consider the following MIMO LTI system with $q=2$, $m=2$, $r=3$ and $n=6$.
\begin{align*}
   x_{t+1}&= \underbrace{\begin{bmatrix}
     4I_2& I_2 &0_2 \\
   0.11I_2 & 0_2 &I_2\\
      0.3I_2 & 0_2 & 0_2 
      \end{bmatrix}}_A x_t+\underbrace{\begin{bmatrix}1 & 0.2& 3 & 0.1& 0.3 & -1 \\2&-5&4&0.9&0.2&-1.1
       \end{bmatrix}^\intercal}_B u_t, \\
       & y_t= \begin{bmatrix} I_2&0_2&0_2\end{bmatrix}x_t \text{, }\;\; x_0=\begin{bmatrix}-1&0.2&1&-3&-2&2.5\end{bmatrix}^\intercal.
\end{align*}
The initial estimates are: $\hat{x}_0=\begin{bmatrix}0&1&2&0.1&0.2&0.3\end{bmatrix}^\intercal$,
\begin{align*}
  \hat{A}_0=  \begin{bmatrix} I_2& I_2 &0_2 \\
   0.51I_2 & 0_2 &I_2\\
      0.6I_2 & 0_2 & 0_2 
    \end{bmatrix} \text{, }
    \hat{B}_0=\begin{bmatrix}0.5 &0.5 &0.5 &0.5&0.5&0.5\\0.5&0.5&0.5&0.5&0.5&0.5
    \end{bmatrix}^\intercal.
\end{align*}
Other parameters are chosen as: $k_0=1000$, $R=I_2$, $k_{min}=0.0001$ and
\begin{align*}
    F=\begin{bmatrix}
    0.4I_2 &  I_2 & 0_2  \\
    0.21I_2 & 0_2 & I_2 \\
    0.2I_2 & 0_2  & 0_2
    \end{bmatrix}.
\end{align*}

\begin{figure}[b]
    \centering
    \includegraphics[scale=0.5]{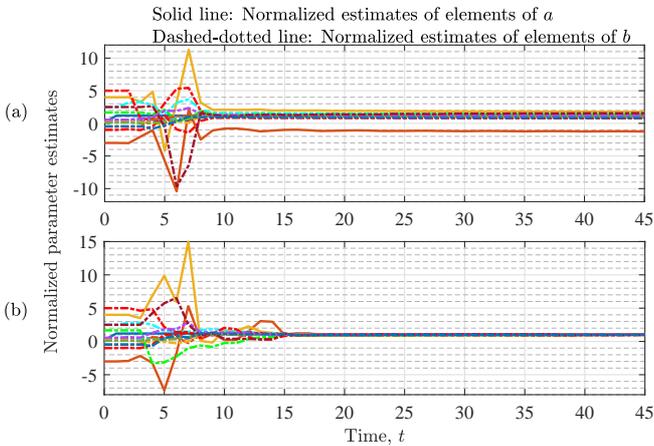}
    \caption{Normalized parameter estimates obtained using (a) non-PE regressor (b) PE regressor.}
    \label{fig2}
\end{figure}

For an LTI system, the regressor is PE if the input is sufficiently rich (\cite{boyd1986necessary}). The plots in Fig. \ref{fig2} (a) and (b) show the normalized parameter estimates in absence and in presence of a sufficiently rich input, respectively. The estimates converge to some finite values in both cases. With a PE regressor (Fig. \ref{fig2} (b)), the normalized parameter estimates converge to $1$. 

\begin{figure}[b]
    \centering
    \includegraphics[scale=0.5]{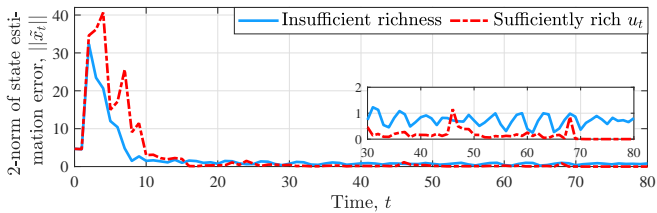}
    \caption{$2$-norm of state estimation error with and without a PE regressor.}
    \label{fig1}
\end{figure}

\begin{figure}[t]
    \centering
    \includegraphics[scale=0.5]{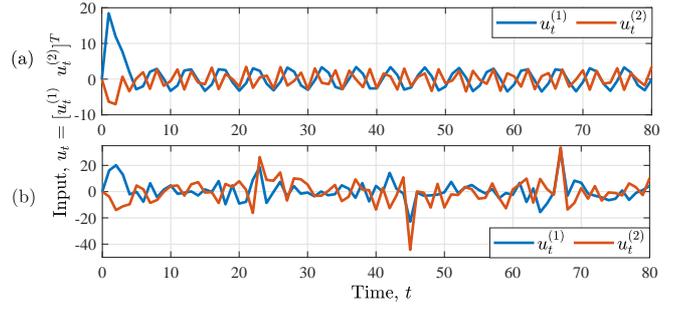}
    \caption{(a) Input without PE (b) Input with PE.}
    \label{fig3}
\end{figure}

The corresponding plots for the $2$-norm of state estimation error are provided in Fig. \ref{fig1}, where the error converges to zero when the input is sufficiently rich. The inputs used to obtain Figs. \ref{fig2} and \ref{fig1} are shown in Fig. \ref{fig3}, where (a) is not sufficiently rich but (b) is. 

The observer proposed in this paper is applicable to discrete-time SISO LTI systems as well. A comparison of the designed observer is made with \cite{suzuki1980discrete} for the following SISO LTI system.
\begin{align*}
   x_{t+1}=& \begin{bmatrix}
    1.52  &1\\-0.6 & 0
      \end{bmatrix} x_t+\begin{bmatrix}0.43\\-0.35
       \end{bmatrix} u_t, \;\; y_t= \begin{bmatrix} 1&0\end{bmatrix}x_t
\end{align*}
with $x_0=0_{2\times 1}$. The initial estimates are: $\hat{x}_0=0_{2\times 1}$,
$  \hat{A}_0=  \begin{bmatrix} 0&1&;&0&0
    \end{bmatrix} \text{, }
    \hat{B}_0=\begin{bmatrix}0&;&0
    \end{bmatrix}.
$
Other parameters are chosen as: $F=\begin{bmatrix}  1.49&1&;&-0.55&0\end{bmatrix}$, $k_0=10000$, $R=1$, $k_{min}=0.0001$ and the weighting factor in \cite{suzuki1980discrete} is $\sqrt{0.5}$.
\begin{figure}[t]
    \centering
    \includegraphics[scale=0.5]{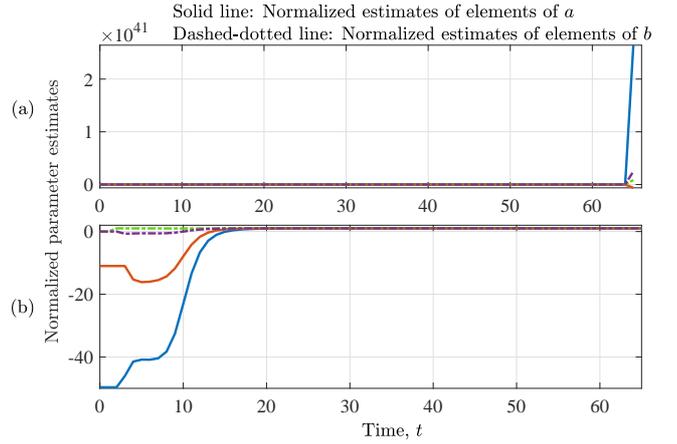}
    \caption{Normalized parameter estimates without using a PE regressor for the observer (a) \cite{suzuki1980discrete} and (b) proposed.}
    \label{fig5}
\end{figure}
\begin{figure}[t]
    \centering
    \includegraphics[scale=0.5]{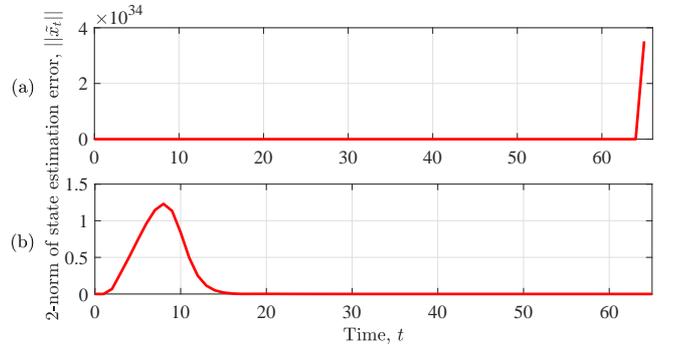}
    \caption{$2$-norm of state estimation error without using a PE regressor for the observer (a) \cite{suzuki1980discrete} and (b) proposed.}
    \label{fig6}
\end{figure}

The example is taken from \cite{suzuki1980discrete} and applied input $u_t=sin(0.2t)$.
Figs. \ref{fig5} and \ref{fig6} show that the parameter and state estimates obtained using \cite{suzuki1980discrete} go unbounded or drift away if the regressor is not PE, whereas the estimates are always bounded with the adaptive observer proposed in this paper.





\section{Conclusion}
A discrete-time MIMO adaptive observer is designed where the parameter learning is separated from the observer dynamics used in state estimation. A covariance resetting based RLS is used for parameter estimation. The state and parameter estimates are shown to be bounded, irrespective of the richness of the input. Presence of a PE regressor (or a sufficiently rich input) further ensures convergence of state and parameter estimation errors to zero. An immediate extension of the work is to ensure robustness to disturbances in the state dynamics and noisy measurements in the output data of \eqref{sys}.


\bibliography{ifacconf}             
                                                   







\appendix
\section{Proof of Lemma \ref{a0}}\label{Append0}
For any vector $\zeta\in\mathbb{R}^{r+mn}$ and any arbitrary matrix $\phi_t$, $\zeta^\intercal\phi_t\phi_t^\intercal\zeta=||\phi_t^\intercal\zeta||^2\geq 0$ and is equal to $0$ when $\zeta=0_{r+mn}$, which proves that $\phi_t\phi_t^\intercal\succeq 0_{r+mn}$.
It is possible to write
\begin{align}
   &\zeta^\intercal\phi_t R^{-1}\phi_t^\intercal\zeta=( \phi_t^\intercal\zeta)^\intercal R^{-1}(\phi_t^\intercal\zeta)\geq \lambda_{min}[R^{-1}](\phi_t^\intercal\zeta)^\intercal(\phi_t^\intercal\zeta) \nonumber\\
   & \Rightarrow \zeta^\intercal\phi_t R^{-1}\phi_t^\intercal\zeta\geq \lambda_{min}[R^{-1}]\zeta^\intercal\phi_t\phi_t^\intercal\zeta \nonumber\\
   & \Rightarrow \zeta^\intercal\phi_t R^{-1}\phi_t^\intercal\zeta\geq \lambda_{min}[R^{-1}] \lambda_{min}[\phi_t\phi_t^\intercal]\zeta^\intercal\zeta \label{zetaphi}
\end{align}
Since $\phi_t\phi_t^\intercal\succeq 0_{r+mn}$ and $R\succ 0_q$, $\lambda_{min}[R^{-1}] \lambda_{min}[\phi_t\phi_t^\intercal]\geq 0$. It implies $\zeta^\intercal\phi_t R^{-1}\phi_t^\intercal\zeta\geq 0$ and is equal to $0$ when $\zeta=0_{r+mn}$, which proves that $\phi_t R^{-1}\phi_t^\intercal\succeq 0_{r+mn}$.

From \eqref{PE}, if $\phi_t$ is PE, then $\lim_{t\rightarrow\infty} \lambda_{min}[\phi_{t}\phi_{t}^\intercal]>0$. Therefore, in the limiting case, $\lambda_{min}[R^{-1}] \lambda_{min}[\phi_t\phi_t^\intercal]> 0$. Substitution of this inequality in \eqref{zetaphi} results in $\lim_{t\rightarrow\infty}\zeta^\intercal\phi_t R^{-1}\phi_t^\intercal\zeta= 0$ if and only if $\zeta=0_{r+mn}$, otherwise $\lim_{t\rightarrow\infty}\zeta^\intercal\phi_t R^{-1}\phi_t^\intercal\zeta> 0$, which proves that $\lim_{t\rightarrow\infty}\phi_t R^{-1}\phi_t^\intercal\succ 0_{r+mn}$.

\section{Proof of Lemma \ref{a1} part (2)}\label{Append1}    

Since $M_{2_{t+1}}\triangleq \Gamma_t^{\frac{1}{2}}\phi_{t+1}R^{-1}\phi_{t+1}^\intercal \Gamma_t^{\frac{1}{2}}$ is symmetric positive semi-definite, it can be decomposed as $M_{2_{t+1}}=M_{3_{t+1}}D_{1_{t+1}}M_{3_{t+1}}^{-1}$, where $M_{3_{t+1}}$ is an orthogonal matrix and $D_{1_{t+1}}$ is a diagonal matrix with the eigenvalues of $M_{2_{t+1}}$ along the diagonal. 

Using this fact and Lemma \ref{a1} (1), the following equation is written.
\begin{align*}
     I_{r+mn}+M_{1_{t+1}}
     = \Gamma_t^{\frac{1}{2}}M_{3_{t+1}}(I_{r+mn}+D_{1_{t+1}})M_{3_{t+1}}^{-1}\Gamma_t^{-\frac{1}{2}}
\end{align*}
The  positive semi-definite matrix $M_{2_{t+1}}$ implies $0\leq \lambda[D_{1_{t+1}}]\leq \infty\Rightarrow 1\leq \lambda[I_{r+mn}+D_{1_{t+1}}]\leq \infty$. Since eigenvalues are preserved under similarity transformation,
\begin{align}
  &\;1\leq \lambda[I_{r+mn}+M_{1_{t+1}}]\leq \infty \label{1inf} \\  \Rightarrow &\;0\leq \lambda\big{[}(I_{r+mn}+M_{1_{t+1}})^{-1}\big{]}\leq 1 \nonumber\\
\Rightarrow& q^\intercal(I_{r+mn}+M_{1_{t+1}})^{-1}q \leq \lambda_{max}\big{[}(I_{r+mn}+M_{1_{t+1}})^{-1}\big{]} q^\intercal q \nonumber\\
  & \hspace{3.4cm} \leq q^\intercal q = q^\intercal I_{r+mn} q\;\;\;\forall q\in\mathbb{R}^{r+mn} \nonumber\\
\therefore &\;(I_{r+mn}+M_{1_{t+1}})^{-1} \preceq I_{r+mn}.\nonumber \hspace{104.4pt}\blacksquare
  \end{align}

\end{document}